\documentclass[12pt,manuscript,authoryear]{aastex}

\usepackage{graphicx}
\usepackage{natbib}
\usepackage{epstopdf}
\usepackage{grffile}

\usepackage{color}
\usepackage{ulem}

\begin{document}

\title{New insight into the Solar System's transition disk phase provided by the unusual meteorite Isheyevo}
  
\author{{\it Short Title: The Solar Transition Disk ~~~~Article Type: ApJL}}

\author{Melissa A. Morris}
\affil{State University of New York, Cortland, NY 13045-0900}
\affil{School of Earth and Space Exploration, Arizona State University, Tempe, AZ 85287-6004}
\email{melissa.morris@cortland.edu}

\and

\author{Laurence A. J. Garvie}
\affil{Center for Meteorite Studies, Arizona State University, Tempe, AZ 85287-6004}

\and

\author{L. Paul Knauth}
\affil{School of Earth and Space Exploration, Arizona State University, Tempe, AZ 85287-6004}

\begin{abstract} 
Many aspects of planet formation are controlled by the amount of gas remaining in the natal protoplanetary disk (PPDs).  Infrared observations show that PPDs undergo a transition stage at several Myr, during which gas densities are reduced.  Our Solar System would have experienced such a stage.  However, there is currently no data that provides insight into this crucial time in our PPD's evolution.  We show that the Isheyevo meteorite contains the first definitive evidence for a transition disk stage in our Solar System.  Isheyevo belongs to a class of metal-rich meteorites whose components have been dated at almost 5 Myr after the first solids in the Solar System, and exhibits unique sedimentary layers that imply formation through gentle sedimentation.  We show that such layering can occur via gentle sweep-up of material found in the impact plume resulting from the collision of two planetesimals. Such sweep-up requires gas densities consistent with observed transition disks (10$^{-12}$ - 10$^{-11}$ g cm$^{-3}$).  As such, Isheyevo presents the first evidence of our own transition disk and provides new constraints on the evolution of our solar nebula.

\end{abstract}


\label{firstpage}

\section{Introduction}
\label{}
The main body of evidence for processes occurring during the early Solar System is found within meteorites.  The combination of this evidence and theoretical modeling has led to a greater understanding of the formation and evolution of the Solar System.  However, several aspects of the evolution of our protoplanetary disk remain unresolved; in particular, the accretion of planetesimals and the formation of planets.  It is generally thought that planet formation depends on the presence of gas within the disk (Kokubo \& Ida 2002; Ikoma \& Genda 2006), but little is known about later densities, during the so-called transition phase.  However, the unusual metal-rich meteorite, Isheyevo, provides insight into this important phase of the early solar nebula.

The metal-rich carbonaceous chondrites (CH: ALH85085-like, CB: Bencubbin-like, and Isheyevo) are intriguing, being mixtures of chondrules, chemically zoned metals, unzoned metal, hydrated lithic clasts, and refractory inclusions, while matrix is largely absent (e.g., Weisberg et al. 2001; Krot et al. 2002; Rubin 2003; Campbell et al. 2005). The zoned metals are hypothesized to have formed from a vapor-melt plume produced during an impact between planetesimals (e.g., Krot et al. 2005; Olsen et al. 2013).  An impact origin is consistent with the majority of components found in these chondrites.  Most chondrules in metal-rich chondrites are cryptocrystalline or skeletal in texture (Weisberg et al.2001; Rubin et al. 2003; Krot et al. 2005; Scott1988; Krot et al. 2001; Hezel et al. 2003), and seem to have formed differently from other chondrules, based upon their inferred thermal histories and their young age.   Whereas the formation of the majority of chondrules has been dated to a range of disk ages 2-3 Myr after the formation of calcium-, aluminum-rich inclusions (CAIs) (Kurahashi et al 2008; Villeneuve et al. 2009), chondrules from CH/CB/Isheyevo chondrites formed almost 3 Myr later (Krot et al. 2005; Krot et al. 2008; Bollard et al. 2013).  Isheyevo contains lithologies characteristic of both CH and CB chondrites \cite{Ivanova2006,Ivanova2008} and phyllosilicate-bearing clasts with extreme $^{15}$N-enrichments \cite{Krot2005,Krot2001,Greshake2001,Briani2009,Bonal2010}, likely of primitive origin (Olsen et al. 2013).  

Most primitive meteorites show evidence of having been compacted, fragmented, and mixed beneath their surfaces, wherein signs of primary accretion are obliterated.  Evidence for accretionary growth of planetesimals is absent in our collection of extraterrestrial materials because such structures would not survive the processes associated with planetary differentiation that gave rise to the iron meteorites and achondrites.  However, Isheyevo preserves primary accretionary structures delineated by a well-sorted mixture of small metal grains, chondrules, CAIs, and clay-bearing matrix lumps \cite{Ivanova2008}. The juxtaposition of fine-grained clasts that experienced extensive aqueous alteration with materials that formed at high temperature shows that Isheyevo is a mechanical mixture of disparate materials.  

\begin{figure}[h]
\centering
\vspace{-0.25in} 
\includegraphics[width=0.30\columnwidth]{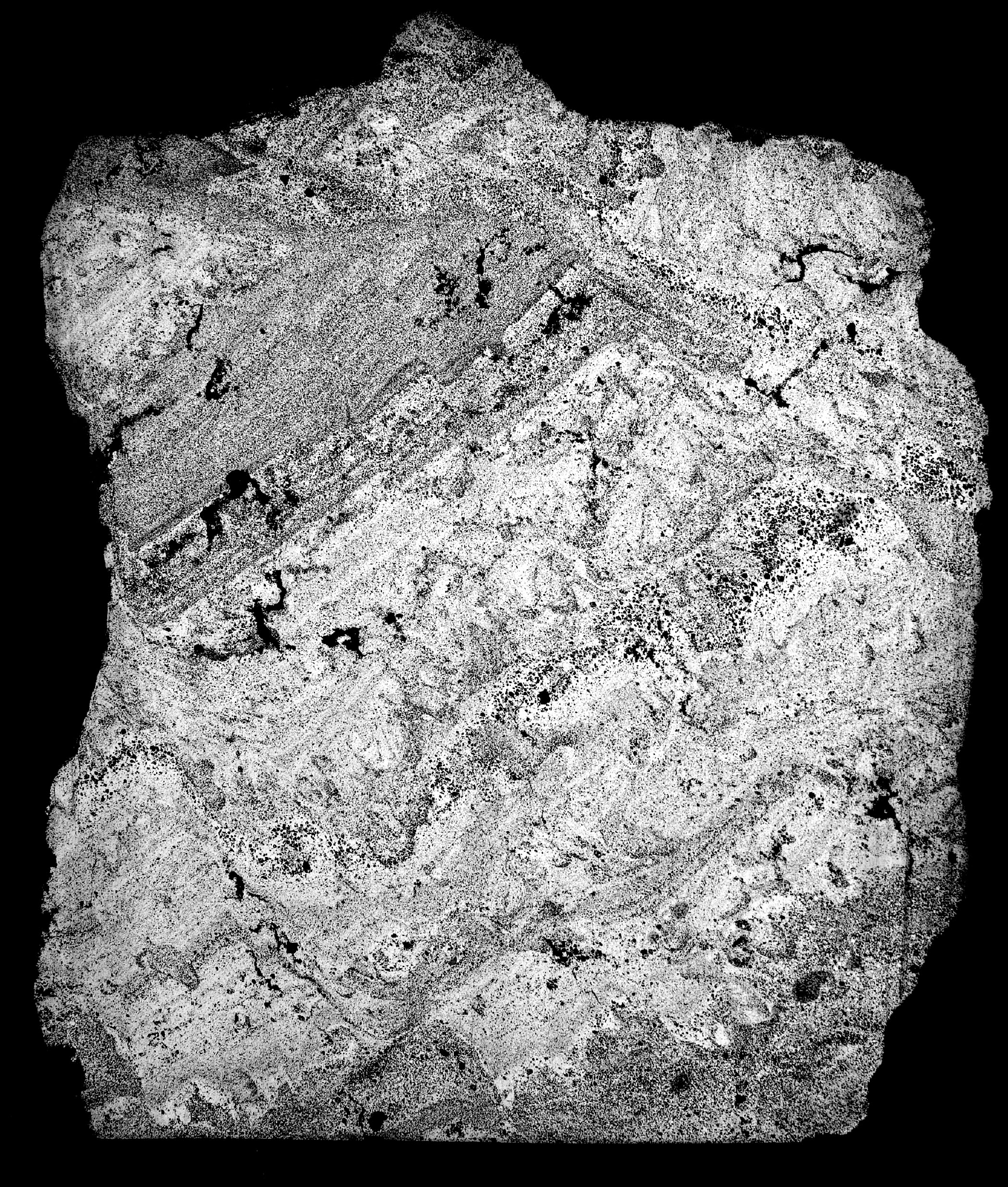}
\includegraphics[width=0.28\columnwidth]{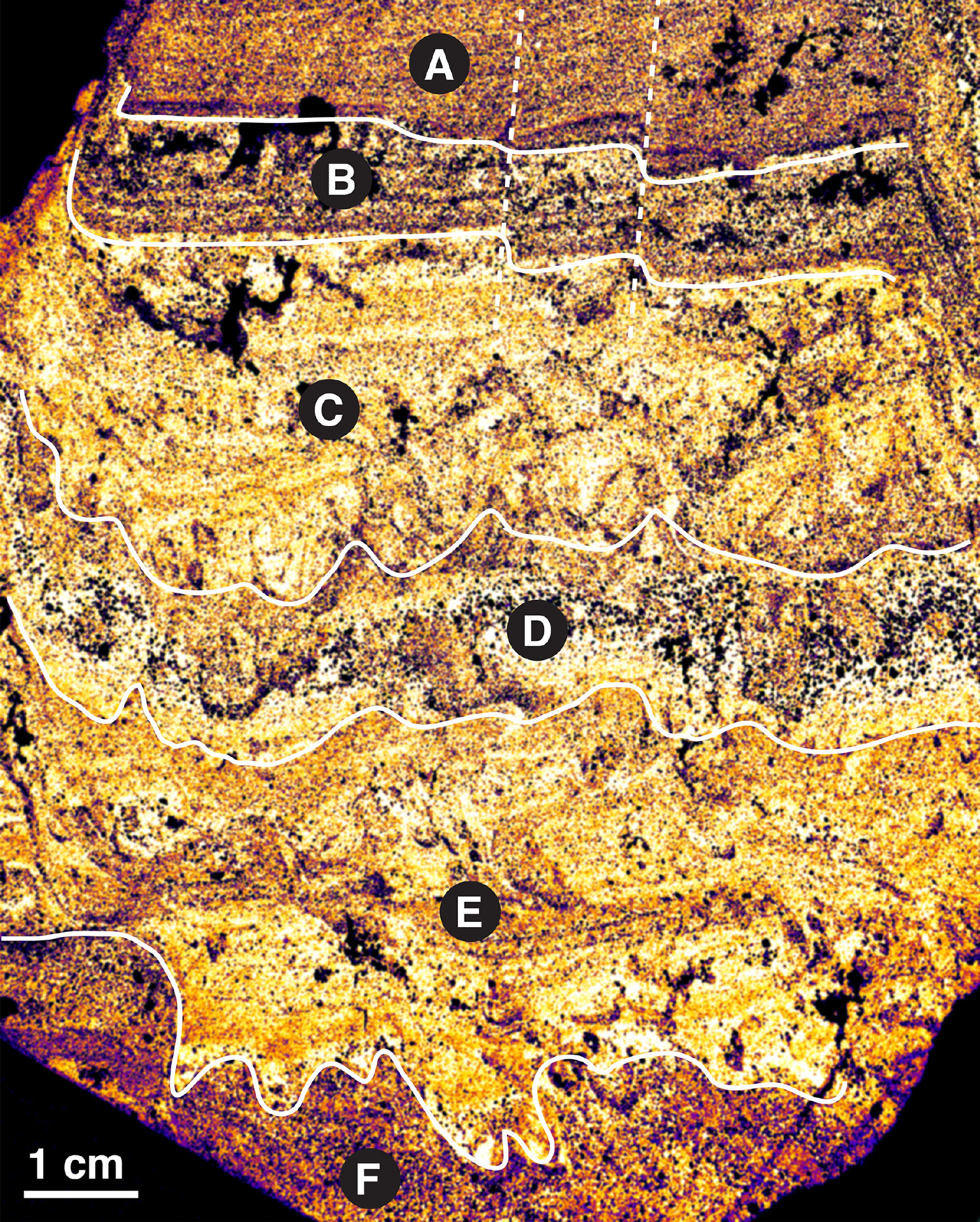}
\includegraphics[width=0.35\columnwidth]{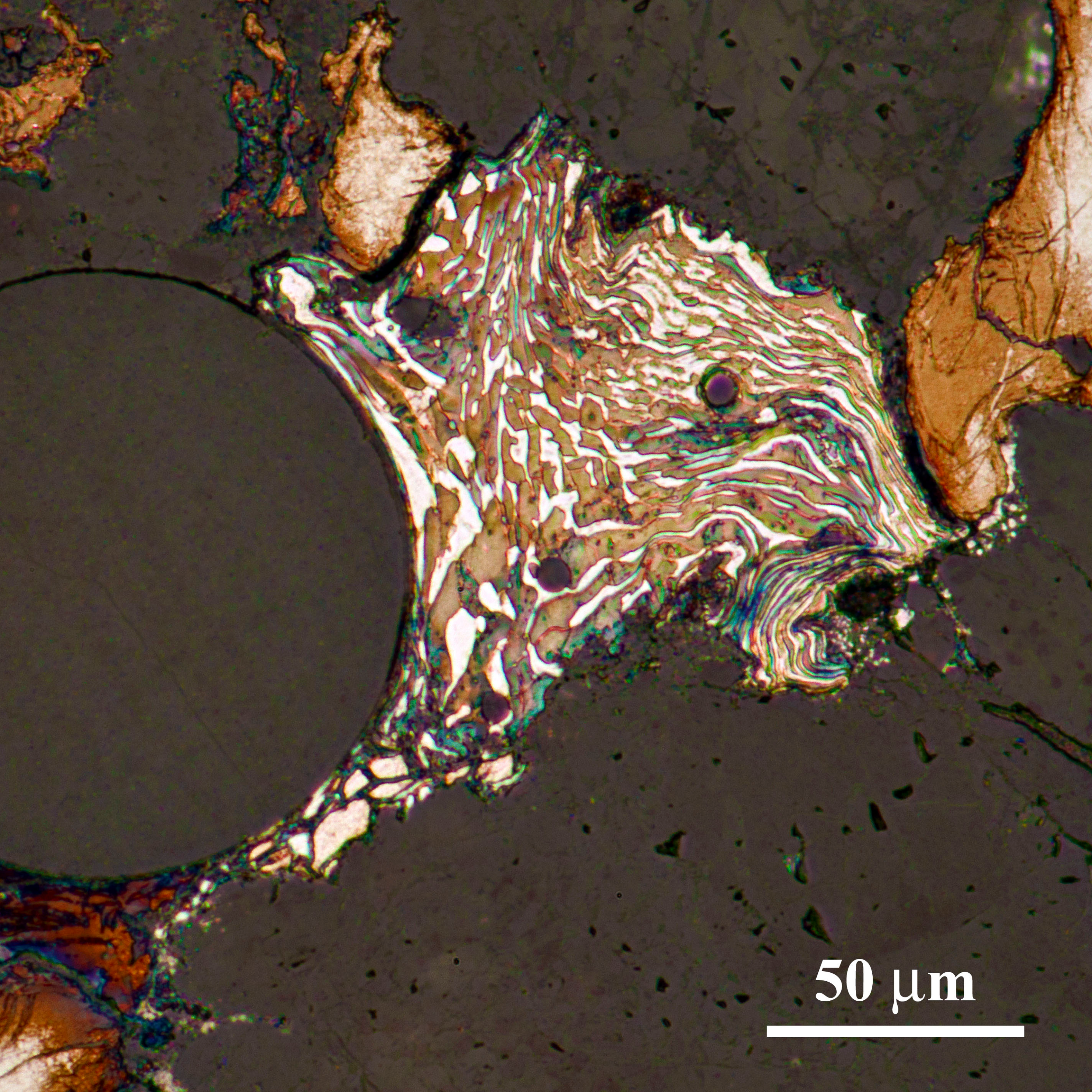} 
\vspace{-0.1in} 
\caption{\footnotesize 
	{\it Left}: Photograph of a 10 x 14 cm slice from Isheyevo. Metal is white and the non-metal components are dark. Evident is the fine laminations near the top of the specimen.  {\it Center}:  False color image of the slice. Whites and yellows represent metal, and the darker colors correspond to the non-metallic components. Layer A is fine-grained with weakly defined laminations and B shows several alternating metal-rich and silicate-rich layers. The base of layer C is silicate rich with lobe-like structures protruding into the coarser-grained layer D. Layer E exhibits several silicate-rich layers and has a largely metal-rich base with well-developed finger-like lobes protruding into layer F. The dashed white lines delineate the faulted sediment.  {\it Right}:  {Reflected-light image showing evidence of cold deformation of a metal particle. The image shows silicates as a uniform dark color - the round structure on the left is a chondrule. The metal has been stained with sodium bisulfite, which stains the metal producing colors that reveal the structure of the metal. The metal particle at the center of the image shows contorted bands of kamacite (stained light brown) and tetrataenite (bright), this contortion is evidence of low temperature deformation. } }
\vspace{-0.08in}
\end{figure}

\section{Observations}

As described in detail in Garvie et al. 2015, Isheyevo preserves primary accretionary structures exemplified by prominent layering and lobe-like structures delineated by the metallic and nonmetallic components.  The laminations are consistent with settling of particles that have not coagulated into aggregates. Layers richer in iron grains protrude downward into layers richer in silicate grains indicating that these are sedimentary load structures akin to those in terrestrial aqueous deposits where dense sediment is deposited over, and protrudes into, a less dense layer \cite{Allen1984}.  The stratigraphic up position is thus established. Also evident are faults that disrupt the planarity of the laminations and can be traced at high angles to the layering. These faults show necking and attenuation of the layers (Figures 1a, 1b), suggesting that the aggregate was weakly cohesive and behaved macroscopically like soft sediment. Microscopic examination shows plastically deformed metal grains (Figure 1c), which together with the clay clasts and chemically zoned metal spheres is evidence of post sedimentary, low-temperature deformation.  These stratigraphic features provide clues to the accretion and formation of the Isheyevo parent body. 

Terrestrial processes that result in sedimentation, such as declining velocity in a fluid flow do not apply in this case.  Therefore, layering in Isheyevo reflects accumulation of material in an accreting environment.  Such layering could occur as a result of high-velocity impacts, but would result in disruption of the layers around larger impacting grains and destruction of clay clasts upon impact, which is not seen.  As such, we investigate other processes that result in gentle sedimentation and layering.

The mixture of chondrules, having textures indicating rapid cooling \cite{Campbell2005}, and metal spheres, which require cooling over days to weeks, (e.g., Goldstein et al. 2007), is consistent with formation in an impact plume (Krot et al. 2005; Olsen et al. 2013).
The implication is that chondrules and metal spheres were mixed with remnants of solid material from the original impactor and then subsequently reaccreted by the surviving planetesimal on relatively short timescales.  Reaccretion via gravitational settling is suggested (e.g., Asphaug et al. 2011).  However, we show that gravitational settling is unlikely, and propose instead that a fan-like sheet of ejecta from the impact was slowed by gas drag and overtaken by the surviving planetesimal at speeds that allowed gentle sedimentation. 

\section{Astrophysical Setting}

Quantitative modeling of glancing blows between molten planetesimals shows that an impact plume, originating primarily from the impactor, can be produced downrange of the collision \cite {Asphaug2011}.  The model predicts the flow of the material and the size of droplets produced \cite{Asphaug2011}.  Chondrule sizes in Isheyevo constrain the sizes of the impacting bodies to be in the tens of km range \cite{Asphaug2011}.  While it has been argued that some of the components in Isyehevo, such as the refractory inclusions and porphyritic chondrules, may have been incorporated from material in the nebula \cite{Krot2008,Krot2009}, such a scenario is inconsistent with our understanding of Solary System evolution, unless they originated as accretionary breccias of impact debris \cite{Krot2014}.  As such, we propose these particular components originate from a solid carapace ($\sim$ 5-km thick) on the impactor \cite{Asphaug2011}.

The results of quantitative modeling \cite{Asphaug2011} show that material ejected from the impact will form a fan-like sheet.  In the case consistent with components from Isheyevo, an $\sim$ 500-km sheet of material is produced downrange a few hours post-collision \cite{Asphaug2011}.  The ejecta includes clasts of material from the unmelted crust and material from the molten interior.  Following impact, the ejecta plume will travel as a unit to a distance that is controlled primarily by how long it takes for the individual meteoritic components to condense.  While the plume remains intact, it is large enough to be unaffected by the nebular gas, so the distance traveled can be calculated based on the initial velocity $\sim V_{\rm esc} \sim 72$ m s$^{-1}$ (Case 5, Asphaug et al. 2011).  At this rate, the leading edge of the sheet of material will reach a distance that is comparable to the Hill radius ($r_{\rm H} = (M_{\rm p} / 3 M_{\odot})^{1/3} \, a$
$\approx 2.1 \times 10^{4} \, {\rm km}$ at 3 AU) in $\sim$ 3.5 days.  After components condense from the ejecta, the fan-like sheet will break apart due to Rayleigh Taylor and Kelvin-Helmholtz instabilities, and will cease to move as a unit.  Chondrules, zoned metal spheres, unzoned metals, and clasts from the original crust of the projectile will then move independently until they are stopped by nebular gas according to their size and material density.  The components are then swept up and reaccreted by the rotating surviving planetesimal downrange from the collision.  This proposed reaccretion scenario depends on the density of nebular gas.  However, the gas density at $\sim$ 5 Myr is poorly constrained.  It is also unclear when our Solar System went through it's transitional stage, during which the amount of gas had decreased substantially from the initial value.  We show that Isheyevo's sedimentary features provide insights into these questions.

\begin{figure}[t]
\centering
\vspace{-0.25in} 
\includegraphics[width=0.46\columnwidth]{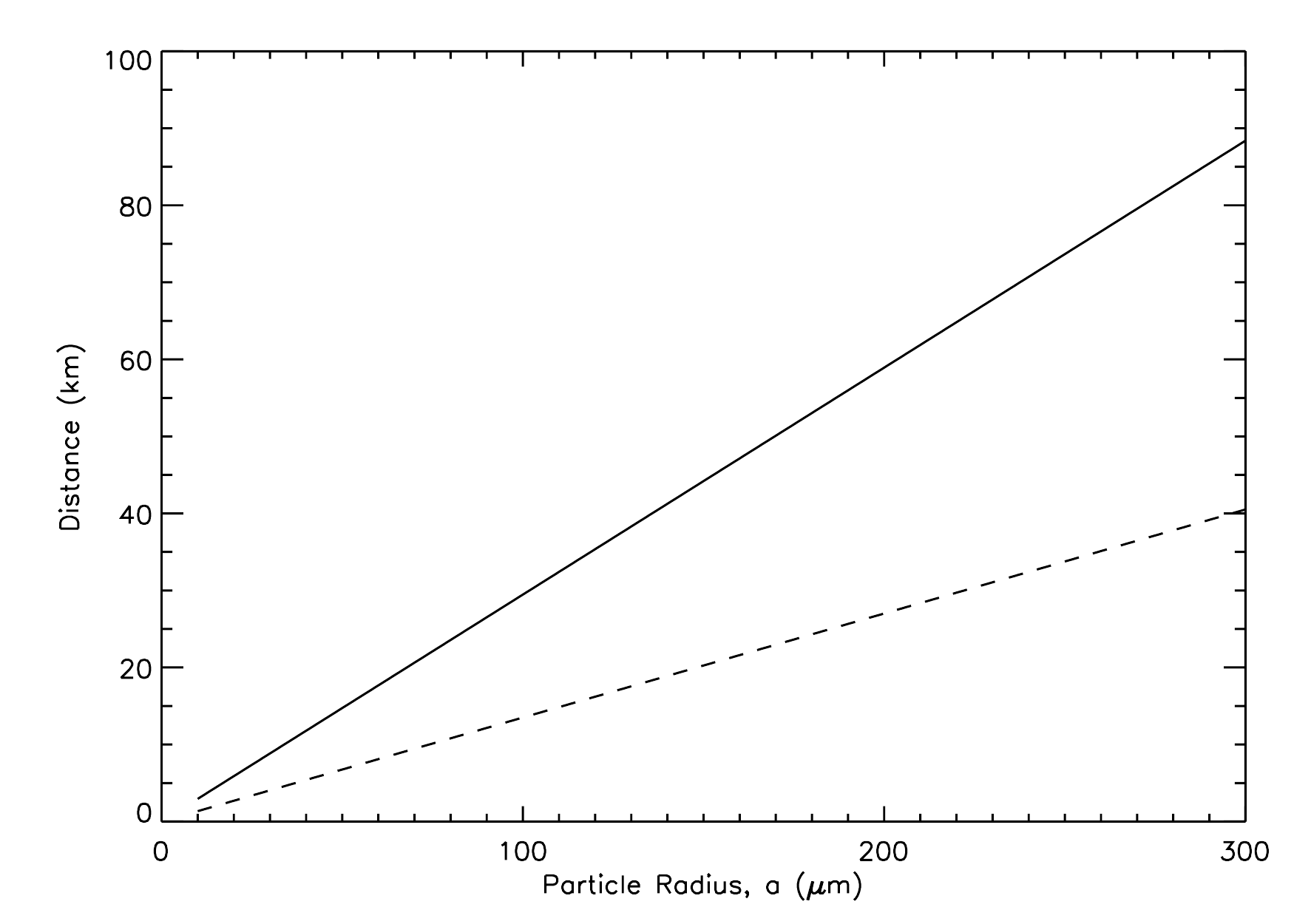} 
\includegraphics[width=0.46\columnwidth]{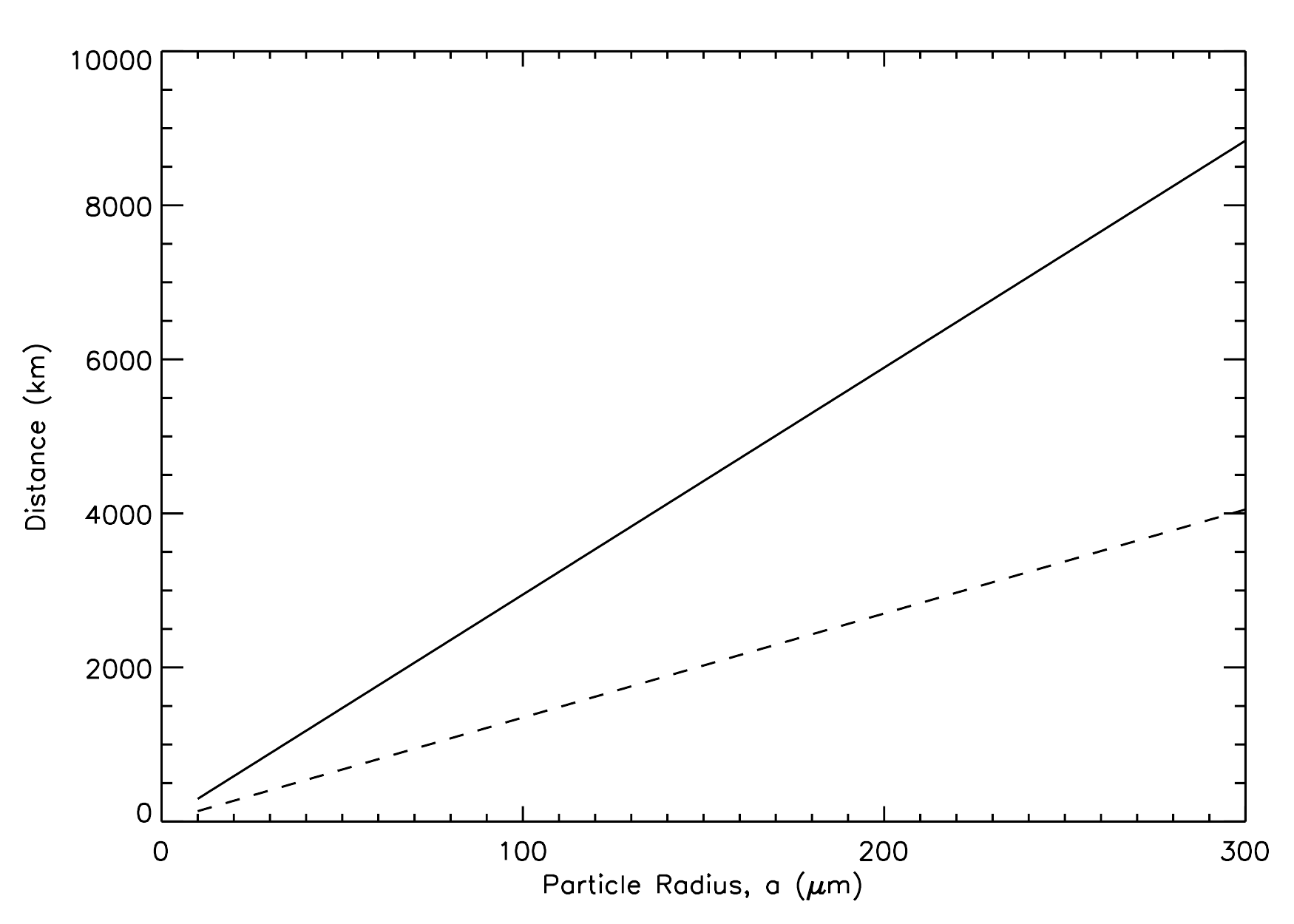} 
\vspace{-0.1in} 
\caption{\footnotesize 
{\it Left}: Distance traveled by metal and silicate spheres of varying radii, ejected from
a common point at speed $V_0 = 30 \, {\rm m} \, {\rm s}^{-1}$, 
before being stopped by gas of density $\rho_{\rm g}  = 10^{-9} \, {\rm g} \, {\rm cm}^{-3}$.
{\it Right}: Same as on the left, except the gas density is $\rho_{\rm g}  = 10^{-11} \, {\rm g} \, {\rm cm}^{-3}$.
}
\vspace{-0.08in}
\end{figure}

\subsection{Aerodynamic sorting of ejecta}

Size sorting of materials in the plume of ejecta occurs because components will travel varying distances before stopping.  The stopping time before spherical droplets of different radii and density recouple to the gas is given by (Cuzzi et al. 2001):
\begin{equation}
t_s=\frac{\rho_s a_s}{c_s \rho_g },
\end{equation}
where $\rho_s$ is the particle density, $a_s$ is the particle radius, $c_s$ is the sound speed at 150 K, and $\rho_g$ is the gas density.  For silicates, we use $\rho_s$  =  3.3 g cm$^{-3}$ (representative of forsterite) and $\rho_s$  = 7.2 g cm$^{-3}$ for Fe-rich metals.  We consider a range of droplet sizes of 10-300 $\mu$m, typical of the sizes of chondrules and metal grains in chondrites.  The distance traveled from a common point of origin by droplets of different radii and material density is given by  $l_{\rm s} \approx V_0 t_{\rm s}$, where $V_0$ is the initial velocity of the droplet.  In gas densities typical of the solar nebula at 2-3 AU at $\sim$ 2 Myr ($\rho_g$ = 10$^{-9}$ g cm$^{-3}$; Desch \& Connolly 2002, Morris \& Desch 2010; Desch et al. 2012), silicate spheres will travel $\sim$ 1-40 km before recoupling to the gas, and metal spheres $\sim$ 3-90 km, (Figure 2, Left).  The dependence of $l_{\rm s}$ on $t_{\rm s}$ results in sorting based on size and composition.  Following the breakup of the impact plume, metal spheres will travel farther than silicates of similar size before their motions are arrested.  In the case of lower gas density ($\rho_g$ = 10$^{-11}$ g cm$^{-3}$), typical of transition disks (Figure 2, Right), metal spherules will travel thousands of kilometers farther than similarly sized silicates.  The difference in stopping times results in the aerodynamical sorting of the particles as shown in Figure 3.

\begin{figure}[h]
\centering
\includegraphics[width=0.60\columnwidth]{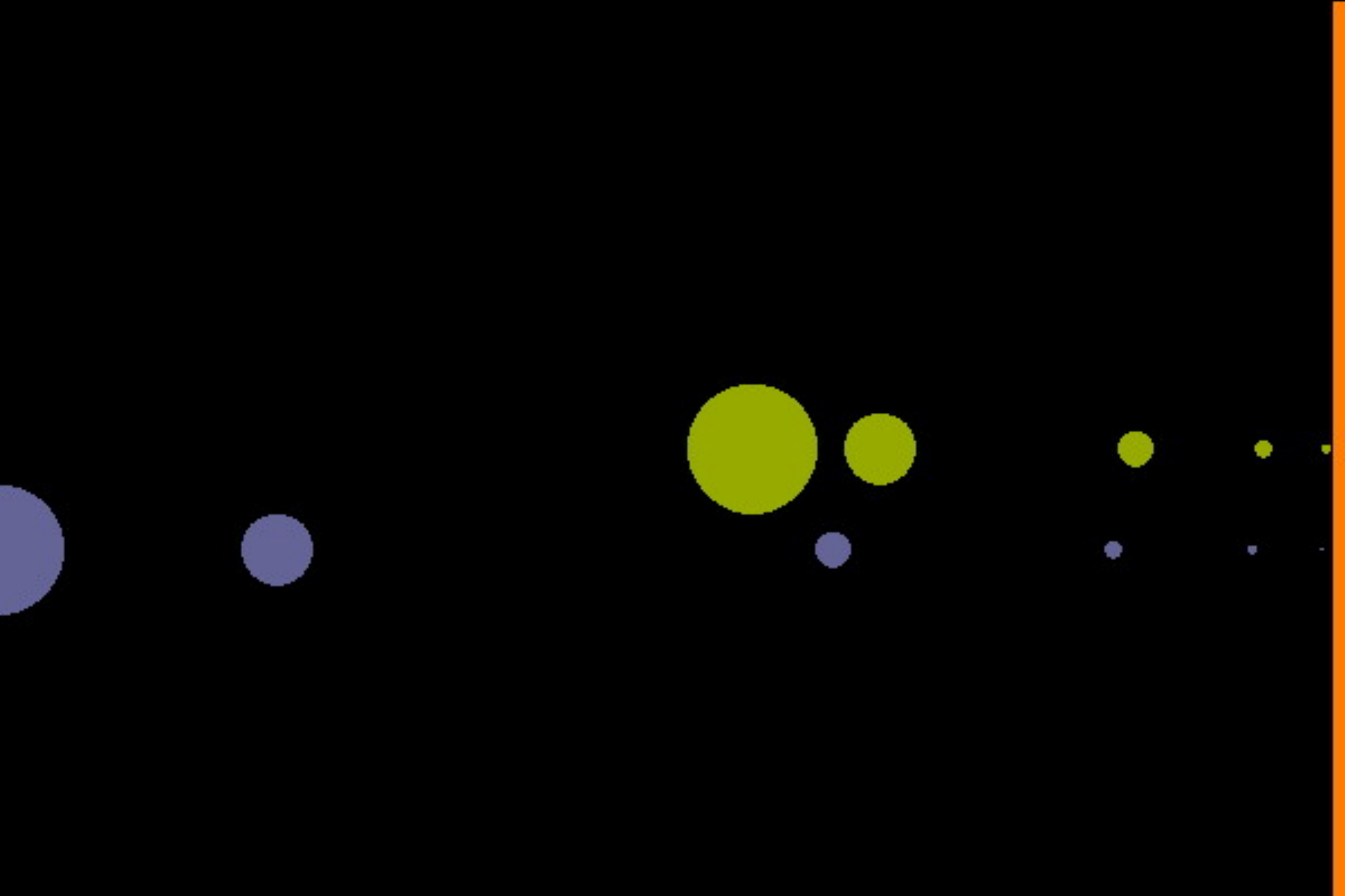} 
\vspace{-0.1in} 
\caption{\footnotesize 
A snapshot showing the distribution of silicate and metal spheres of different radii, originating from a common point, after becoming recoupled to the gas.  Silicate spheres are shown in green and metal spheres are shown in blue.  The Isheyevo parent body (indicated by the thin orange area) will sweep up particles as it travels from right to left.
}
\end{figure}

After silicate and metal particles are stopped and recouple to the nebular gas, they must then reaccrete onto the Isheyevo parent body; otherwise the components would disperse into the nebula.  Previously proposed scenarios for reaccretion have invoked gravitational settling \cite{Krot2008,Asphaug2011}, but we argue that this process is too slow to present a reasonable method for reaccretion.  

\subsection{Timescales for gravitational settling}
The gravitational settling time is given by $z/v_t$, where $z$ is the distance from the target, and the terminal velocity $v_t$= $gt_{\rm s}$, where $g$ is the local gravity.  In a nebula with $\rho_g$ = 10$^{-11}$ g cm$^{-3}$, the time for silicate particles of 10-300 $\mu$m to settle from the Hill sphere of a planetesimal 70 km in diameter is 83 - 2500 yr.  For similarly sized metal particles, the settling time is 38 - 1150 yr.    Over such long time periods, particles would disperse through the nebula due to turbulence in the gas long before they are able to gravitationally settle to the body.  Even were it possible for the particles to settle before dispersion, layering would not occur.  As a consequence of the long timescales involved for gravitational settling, particles would reach terminal velocity long before nearing the body, reaching the surface simultaneously and erasing the effects of the size sorting.  For higher gas densities,  gravitational settling is even more improbable.  For example, at $\rho_g$ = 10$^{-9}$ g cm$^{-3}$, the time for particles to settle from the Hill sphere are two orders of magnitude higher than for $\rho_g$ = 10$^{-11}$ g cm$^{-3}$.  Therefore, reaccretion by gravitational settling is implausible.

\subsection{Reaccretion via ``sweep-up"}

Our calculations and data from Isheyevo are consistent with the rotating impacted body sweeping up material downrange of the collision, as it continues on with a slight velocity relative to the gas and particles.  This sweep-up scenario is compatible with the size sorting of silicates and metals in Isheyevo.  Our determination of the distribution of particles (Figure 3) predicts that chondrules will be swept up by the parent body with smaller-sized metals.  In general, we observe that this is the case within Isheyevo.  Size measurements from a representative piece of Isheyevo show that metals have a radius of 33 $\mu$m (n=161) and silicates, including clay clasts, have a radius of 60 $\mu$m (n=56).  In addition, sweep-up at low velocity is necessary to preserve the clay-rich clasts.  

Our calculations suggest that sufficient mass can be swept up to produce meters-thick layers of particles resembling those found in Isheyevo.  The fraction of particles swept up by a parent body with diameter $D \approx 70 \, {\rm km}$
will be $f = (\pi D^2)/(4 A)$, where $A$ is the area of the fan-like sheet of material at the time it is reaccreted.
If we assume the sheet of material stops moving as a unit after spreading to a distance 
$r \sim$ Hill radius, and assuming homologous expansion of a 500 km x 500 km square at 3.3 hours, the sheet has area $\approx 2 \times 10^8 \, {\rm km}^2$,  so that 
$f \sim 2 \times 10^{-5} $. 
Gravitational focusing can increase this by a factor $\sim (v_{\rm esc}/2V)^2$, where
$V \sim 1 \, {\rm m} \, {\rm s}^{-1}$ may reflect the random velocities of particles, 
yielding $f \sim 3 \times 10^{-2}$. 
In gas of density $\rho_{\rm g} = 10^{-11} \, {\rm g} \, {\rm cm}^{-3}$, and for an ejected mass $3 \times 10^{19} \, {\rm g}$ \cite{Asphaug2011}, the mass of solids reaccreted 
is $\sim 8 \times 10^{17} \, {\rm g}$. 
This mass of solids is sufficient to coat the entire asteroid surface to a depth of about 1 meter, 
or cover a fraction of the asteroid surface to greater depth, depending on its rotation rate. 
These preliminary calculations demonstrate that for the stopping lengths we consider typical,
it is possible for the parent body to sweep up sufficient mass to produce the CH/CB/Isheyevo
chondrites, provided the asteroid moves in the same direction as the ejecta, the initial velocity of the sheet of material, $V_0$, is low 
($\sim V_{\rm esc}$), and the gas density is high enough to arrest the motions of the 
particles before they travel much farther than $\sim 2 \times 10^{4}$ km.  

\section{Astrophysical Implications}

Reaccreation by sweep-up at low velocity, while retaining evidence of the aerodynamic sorting, places constraints on the the density of nebular gas and the degree of turbulence in the nebula.  

\subsection{Nebular gas density and turbulence needed to meet meteoritic constraints}

Zoned metal particles, such as those found in Isheyevo, are interpreted to have formed and cooled in a matter of days to weeks, based on their chemical zoning profiles \cite{Goldstein2007,Meibom2000,Petaev2001,Petaev2003,Campbell2004}. This time constraint imposes bounds on the gas density of the protoplanetary disk during their formation.  Metal spheres formed in impacts and dispersed in gas at densities typical of the formation of most chondrules at around 2 Myr post CAIs ($\rho_g$ = 10$^{-9}$ g cm$^{-3}$) would be stopped and reaccreated on timescales too short to allow for their condensation.  In order to meet the constraints on cooling times for metal spheres, sweep-up must occur within days to weeks.  According to our parameter studies (Table 1), gas densities in the range of  $\rho_g$ = 10$^{-11}$ to 10$^{-12}$ g cm$^{-3}$ are indicated for accreting bodies moving with a sweep-up velocity of $v_{\rm su}$ = 25 - 500 m s$^{-1}$.   Gas densities lower than $\rho_g$ = 10$^{-12}$ g cm$^{-3}$ would result in longer stopping times, causing dispersal of the impact products into the nebula, without reaccretion.  


In order to preserve the aerodynamic sorting effects of  the particles, they must be swept up before they are mixed by turbulence.   We have employed the methods described by  \cite{Cuzzi2004} to determine the mixing timescale for individual components, once they have recoupled to the gas.  The effective viscosity in a weakly turbulent nebula is given by $\nu_{\rm t} = \alpha c_{\rm s} H$, where $\alpha$ is a dimensionless parameter determined by the mass accretion rate of the nebula, $c_{\rm s}$ is the sound speed, and $H$ is the scale height of the disk.  Both models and observations suggest that typically $\alpha \sim 10^{-5}$ to $10^{-2}$ and $H \sim R/20$, where $R$ is the distance from the central star \cite{Cuzzi2004}.   The diffusivity due to turbulence is $D = \nu_{\rm t}/Pr_{\rm t}$, where $Pr_{\rm t}$ is the Prandtl number, typically assumed to be $Pr_{\rm t}$ = 1, giving $D = \nu_{\rm t}$ \cite{Cuzzi2004}.  The timescale for mixing of particles separated by a distance $L$ is then $t_{\rm mix} = L^2/D$.  

Given $\alpha = 10^{-5}$, and a maximum particle separation distance $L_{\rm max} \sim$ 8800 km (as shown in Figure 2),  $t_{\rm mix} \sim$ 100 hours.  The largest components must be swept up within this time frame in order to preserve the sorting depicted in Figure 3, which requires that the Isheyevo parent body must  move at a velocity $\geq$ 24 m s$^{-1}$ relative to the particles.  This is a lower limit to the relative velocity, $V_{\rm rel}$, we expect as a result of the collision.  
 At minimum particle separation of $\sim$ 13 km, applicable to smaller particles, mixing can occur within one second.  The average size of particles in Isheyevo indicate separation distances of $L \sim$ 6950 km, so $t_{\rm mix} \sim$ 70 hours, which requires that the body move at $V_{\rm rel} \sim $  28 m s$^{-1}$.
An upper limit to the relative velocity is provided by the preservation of the integrity, both thermally and mechanically, of the clay-like clasts.  It is therefore likely that the sweep-up velocity falls at the lower limit of the range indicated by particle size.  
This relative velocity, combined with the constraints on cooling rates of metal particles, is consistent with gas densities of $\rho_g$ = 10$^{-11}$ to 10$^{-12}$ g cm$^{-3}$ (Table 1). It is also important to note that our calculations of the mixing timescale (combined with the size sorting observed in Isheyevo) place constraints on the degree of turbulence in the disk, since for larger $\alpha$, $t_{\rm mix}$ is correspondingly shorter.  We find that for $\alpha > 10^{-5}$, preservation of sorting, such as that observed in Isheyevo, is unlikely.  

\begin{deluxetable}{cccccccc}
\tablecolumns{6}
\small
\tablewidth{0pt}
\tablecaption{Results of Parameter Study on Sweep-up Time}
\tablehead{
\colhead{$v_{\rm su}$}&
\colhead{10$^{-09}$ \tablenotemark{a}}&
\colhead{10$^{-10}$} & 
\colhead{10$^{-11}$}& 
\colhead{10$^{-12}$}&
\colhead{10$^{-13}$}
 }
\startdata
25 m s$^{-1}$&$<$1.0 \tablenotemark{b}& 0.3-10&3-98& 33-982&327-9821 \\
50 m s$^{-1}$&$<$0.5 & 0.2-5&2-49& 20-491&200-4910 \\
100 m s$^{-1}$ &$<$0.2 & 0.08-2&0.8-24& 8-245&80-2455 \\
250 m s$^{-1}$&$<$0.01 & 0.03-0.1&0.3-10& 3-98&30-982 \\
500 m s$^{-1}$&$<$0.05 & 0.01-0.5&1-5& 2-50&16-500 \\
\enddata
\tablenotetext{a}{Gas density in units of g cm$^{-3}$.}
\tablenotetext{b}{Sweep-up in units of days.}
\label{table:par}
\end{deluxetable}

\section{Conclusion}

Our calculations show that the components found in the Isheyevo meteorite are consistent with sweep-up at low velocity within gas of density $\rho_g$ = 10$^{-11}$ to 10$^{-12}$ g cm$^{-3}$.  These densities are consistent with those of observed transition disks \cite{Salyk2009,Williams2011}.  Through Isheyevo's association with meteorites that have components dated at around 5 Myr (Krot et al. 2005; Ivanova et al. 2006; Bollard et al. 2013), we infer that this important stage in the evolution of the Solar System occurred at $\sim$ 5 Myr.  We also show that in order to preserve the observed sorting of Isheyevo components, sweep-up must occur in a turbulent disk with $\alpha < 10^{-5}$.
We therefore conclude that Isheyevo, the oldest known sedimentary rock, accreted in the Solar System's mildly turbulent transition disk, a heretofore purely theoretical phase in the primordial solar nebula.  





\section{Acknowledgements}
We wish to thank Bill Bottke and Jeff Cuzzi for their thoughtful comments and suggestions.  M.A.M. was supported by NASA Cosmochemistry grant NNX14AN58G.  L.A.J.G was supported by NASA Origins of Solar System grant NNX11AK58G.

\label{lastpage}

\end{document}